\documentstyle[emulateapj,psfig,epsfig]{article}

\begin{document}

\received{}
\revised{}
\accepted{}

\lefthead{M. Catelan}
\righthead{}

\slugcomment{The Astrophysical Journal, in press}

\singlespace

\title{Horizontal-Branch Models and the Second-Parameter Effect.\\  
       III. The Impact of Mass Loss on the Red Giant Branch,\\ 
       and the Case of M5 and Palomar~4/Eridanus}

\author{M.~Catelan\altaffilmark{1,}\altaffilmark{2,}\altaffilmark{3}}
\affil{NASA Goddard Space Flight Center,
       Code 681, 
       Greenbelt, MD 20771, USA}
\altaffiltext{1}{Hubble Fellow.}
\altaffiltext{2}{Visiting Scientist, Universities Space Research
   Association.}      
\altaffiltext{3}{Current address: University of Virginia, Department 
of Astronomy, P.O.~Box~3818, Charlottesville, VA 22901-2436; 
e-mail: catelan@virginia.edu.}

\begin{abstract}
Deep {\em Hubble Space Telescope} (HST) photometry has recently 
been presented for the outer-halo globular clusters 
Palomar~4 and Eridanus. The new high-precision color-magnitude 
diagrams obtained for these globulars have allowed a measurement
of their ages relative to M5 (NGC~5904), which is a well-observed, 
much closer cluster. Assuming that the globular clusters share 
the same chemical composition, Pal~4/Eridanus have been reported 
to be younger than M5 by $\approx 1 - 2$~Gyr, based on both the 
magnitude difference between the horizontal branch (HB) and the 
turnoff and the difference in color between the turnoff and the 
lower subgiant branch. In the present article, we address the 
following question: What age difference would be 
required to account for the difference in HB 
types between M5 and Pal~4/Eridanus, assuming age to be the 
``second parameter"? We find that, unless all these clusters 
(including M5) are younger than 10~Gyr, such an age difference 
is substantially larger than that based on an analysis of the 
cluster turnoffs. To reach such a conclusion, six different 
analytical mass loss rate formulae (reported in an Appendix),  
all implying a dependence of mass loss on the red giant branch 
on age, were employed. Our results appear to be in conflict 
with claims that age can be the only second parameter in the 
Galactic globular cluster system. 
\end{abstract}


\section{Introduction}
One of the most important ingredients for the construction of a 
model of the formation of the Galaxy concerns whether 
the globular clusters (GCs) in the Galactic outer halo are 
younger or older than those in the inner halo (e.g., Mironov
\& Samus 1974; Searle \& Zinn 1978; Zinn 1980, 1993; van den 
Bergh 1993; Majewski 1994).

The outer halo of the Galaxy is not well populated. From 
Table~7 in Borissova et al. (1997), one finds that in the 
``extreme" outer halo (galactocentric distances 
$R_{\rm GC} > 50$~kpc) there are five very scarcely populated 
and loose clusters with (mostly) red horizontal-branch (HB) 
morphologies (Palomar~3, Pal~4, Pal~14, Eridanus, and AM-1), 
and one cluster with a blue HB. This blue-HB 
globular---NGC~2419---is, however, more massive (by a factor 
of $\simeq 8$) than the sum of {\em all} outer-halo GCs with 
red HBs. {\it Hubble Space Telescope} (HST) observations have 
revealed that NGC~2419 is coeval with M92 (NGC~6341) (Harris 
et al. 1997), a much closer blue-HB globular which has always 
been considered to be among the very oldest GCs in the Galaxy 
(e.g., Bolte \& Hogan 1995; Pont et al. 1998; Salaris, 
Degl'Innocenti, \& Weiss 1997; VandenBerg, Bolte, \& Stetson 
1996). 

Until recently, however, little information was available on the
ages of the remaining extreme outer-halo GCs. HST observations
have again helped remedy the situation. Stetson et al. (1999) 
have presented additional results of their ongoing HST survey of 
GCs lying at $R_{\rm GC} > 50$~kpc. In particular, they presented 
deep WFPC2 F555W, F555W--F814W ($V$, $V-I$) 
color-magnitude diagrams (CMDs) for Pal~4 and Eridanus---both of
which have exclusively red HBs.\footnote{Stetson et al. (1999) also 
presented HST observations for Pal~3 and derived its age relative  
to M3 (NGC~5272). We defer analysis of this pair to a future paper
because of the current difficulty in determining ``representative" 
HB morphology parameters for M3, a cluster which appears to show 
a strong radial gradient in HB type. The latter conclusion can be 
obtained from a comparison among the datasets presented by Buonanno 
et al. (1994), Ferraro et al. (1997a) and Ferraro (1998)---the 
latter referring to HST-WFPC2 data for the innermost cluster regions 
(Ferraro et al. 1997b).}  
Stetson et al. undertook an analysis of the ages of these GCs, as 
provided by traditionally employed techniques 
(e.g., Stetson, VandenBerg, \& Bolte 1996 and references therein), 
and found that, under the assumption that M5 (NGC~5904)
and Pal~4/Eridanus share the same chemical composition, 
these extreme outer-halo GCs with red HBs 
are younger than M5 ($R_{\rm GC} \simeq 6.2$~kpc; Harris 1996) by 
$\approx 1.5 - 2$~Gyr. 

VandenBerg (1999a) has recently reanalyzed the HST CMDs for the 
extreme outer-halo GCs with red HBs. Again assuming identical 
chemical compositions, he supports slightly smaller age differences 
($\approx 1 - 1.5$~Gyr) between M5 and Pal~4/Eridanus than reported 
by Stetson et al. (1999). Therefore, 2~Gyr appears to be a safe 
upper limit on such an age difference.

\parbox{3in}{\epsfxsize=3.25in \epsfysize=3.0in \epsfbox{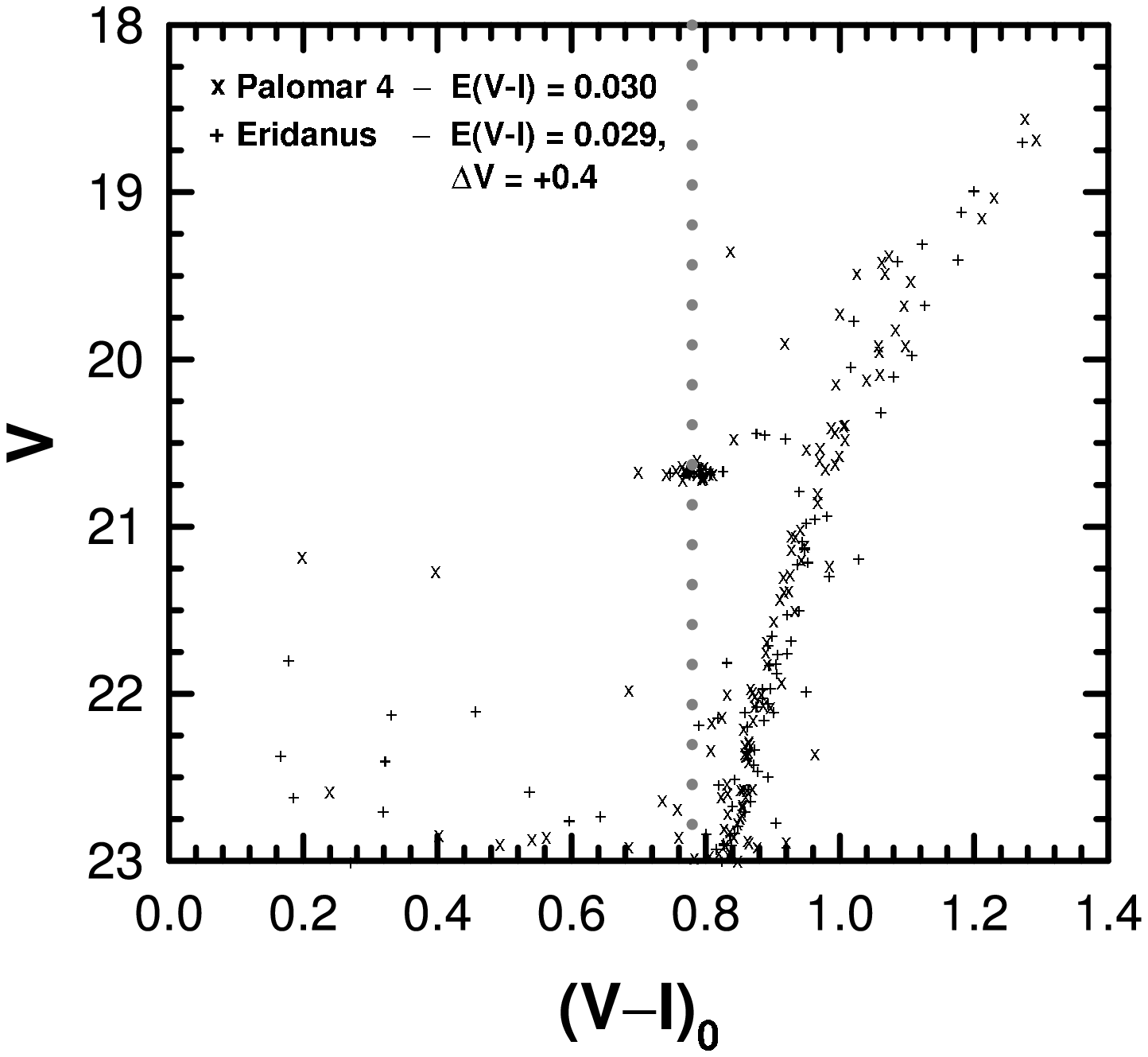}} 
\vskip 0.1in
\centerline{\parbox{3.5in}{\footnotesize  {\sc Fig.~1.---}
Combined HST CMDs for Pal~4 ($\times$) and Eridanus (+).  
          As indicated, reddenings of $E(V-I) = 0.030$ and 0.029~mag 
          have been assumed for Pal~4 and Eridanus, respectively
          (based on Schlegel et al. 1998).  
          The Eridanus CMD has been shifted by 
          $\Delta V = +0.4$~mag, thus accounting for the relative 
          distance moduli of the two clusters (VandenBerg 1999a). 
          The vertical dotted line, in gray, indicates the mean 
          color of the bulk of the HB populations in the two 
          clusters, $\langle (V-I)_0 \rangle = 0.78$~mag. Note 
          that the Schlegel et al. reddening values imply 
          intrinsically bluer HBs than do the canonical reddening  
          values tabulated by Harris (1996). 
 }}
\vskip 0.25in

Though a lower age for Pal~4 and Eridanus would {\em qualitatively}
appear consistent with their red HB types, Stetson et al. (1999) did 
not attempt to provide a reliable {\it quantitative} description of 
how large an age difference would be required to explain the 
difference 
in HB morphology between Pal~4/Eridanus and M5. Lee, Demarque, \& 
Zinn (1994) have recently stated: ``only a small number of
clusters have been dated to sufficiently high precision to test the
hypothesis that the second parameter is age, and there is some doubt
that the detected age differences are consistent with the HB
morphologies of the clusters. If they are not, this would suggest
that age cannot be the sole second parameter." We concur with such
a statement and emphasize, therefore, that tests of age as the 
second parameter cannot be properly carried out without the 
required comparison with adequate models of the HB morphology of 
the clusters under consideration. 

As we have done in the previous papers of this series (Catelan \& 
de Freitas Pacheco 1993, 1994, 1995), we shall provide here the
quantitative estimates of the age difference that 
is required to explain the HB morphologies of M5 vs. 
Pal~4/Eridanus. {\it We shall assume that age is the 
sole second parameter.} Using results reported in an Appendix  
(see also Catelan 1999), we shall
examine in detail the effect of an age-dependent red giant branch
(RGB) mass loss upon the inferred age differences, since there have
been suggestions (e.g., Lee et al. 1994) that such an age dependence 
may help explain the second parameter phenomenon in terms of age.

We begin in the next section by describing the observational 
data for M5, Pal~4 and Eridanus employed in the present study. 
In \S3, we describe our technique for obtaining synthetic HB 
models for these clusters. In \S4, we explain how the age difference 
between Pal~4/Eridanus and M5 that is required to account for their 
different HB types was obtained from the models, taking into 
account several different analytical mass loss formulae for the 
mass loss in red giants. Finally, we present conclusions and provide 
additional discussion in \S5.

\section{Observational Data}

\subsection{HB Morphology of M5}
Sandquist et al. (1996) have provided a very extensive account of
the CMD morphology of M5. Recently, Sandquist (1998) has kindly
readdressed the HB morphology parameters for the cluster. His
latest values can be found in Table~1. 
In column~1, the Mironov (1972) index $B/(B+R)$ is given.
In column~2, one finds the so-called ``Lee--Zinn parameter" 
$(B-R)/(B+V+R)$, first defined and used by Zinn (1986). In column~3, 
Buonanno's (1993) index $(B2-R)/(B+V+R)$, where $B2$ is the 
number of blue-HB stars bluer than $(\bv)_0 = -0.02$~mag, is 
provided. As usual, $B$, $V$, $R$ are the numbers of blue, variable 
(RR Lyrae--type) and red HB stars, respectively. The final two
columns provide two alternative values for Fusi Pecci's $HB_{\rm RE}$ 
indicator (a ``subjective" estimate of the red end of the HB
distribution in \bv; Fusi Pecci et al. 1993), where the first 
disregards the presence of a few red-HB stars lying ``above 
the zero-age HB," and the second takes such stars into account 
(Sandquist 1998). For additional information and references related 
to these indices, the reader is referred to Catelan et al. (1998).

\vskip 0.25in
\parbox{3in}{\epsfxsize=3.25in \epsfysize=0.85in \epsfbox{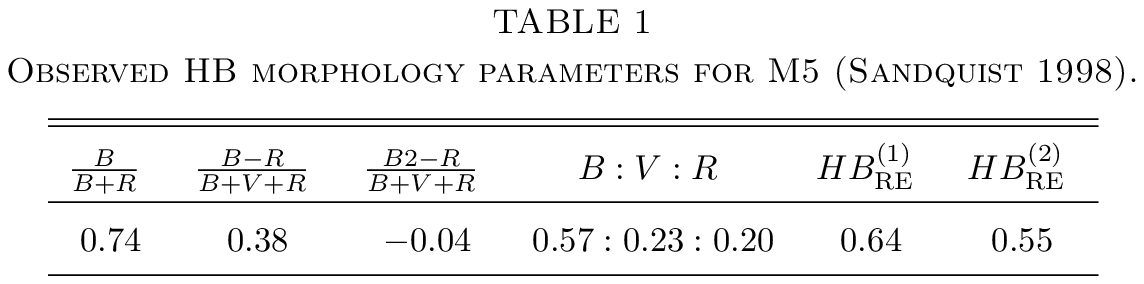}} 
\vskip 0.25in

It is important to note that Buonanno's (1993) index is 
(unfortunately) reddening-dependent. For M5, 
$E(\bv) = 0.03\pm 0.01$~mag (cf. Sandquist et al. 1996 and 
references therein). The value of the Buonanno parameter provided 
in Table~1 corresponds to the assumption that $E(\bv) = 0.03$~mag. 
Sandquist (1998) has kindly evaluated the effect of the reddening 
uncertainty upon this index for M5: he finds that, if the 
reddening is actually 0.02 or 0.04~mag, then this ratio would have 
the values $-0.056$ or $-0.016$, respectively.  

\subsection{HB Morphology of Palomar~4 and Eridanus}
Pal~4 and Eridanus have exclusively red HBs, making the computation 
of most of the above HB morphology indices meaningless for our 
purposes.  

From the HST CMDs (see Stetson et al. 1999), 
it is clear that both Pal~4 and Eridanus share very similar HB 
morphologies. A combined CMD for the two clusters is shown in 
Figure~1. The stars were de-reddened by the indicated amounts, 
based on Schlegel, Finkbeiner, \& Davis (1998). The Eridanus 
data were further shifted by $\Delta V = +0.4$~mag, in order to 
account for the relative distance moduli of the two clusters 
(VandenBerg 1999a). As indicated by the vertical dotted line, 
the HB color distribution clearly shows a peak at 
$(V-I)_0 \simeq 0.78$~mag (and little scatter around this
point). Eridanus has a few ($\sim 3$) stars scattered towards 
brighter magnitudes and redder colors than does Pal~4; however, 
as shown below, this feature can be accounted for by statistical 
fluctuations related to the small number of HB stars (a total of 
$\approx 25$) available in the HST samples and evolution away 
from the zero-age HB (ZAHB).

\section{Theoretical Framework: Synthetic HBs}
The HB evolutionary tracks employed in the present project are
the same as described in Catelan et al. (1998). The following
chemical composition was assumed: main-sequence helium
abundance $Y_{\rm MS} = 0.23$, overall metallicity $Z = 0.001$ 
(see Sneden et al. 1992; Sandquist et al. 1996; Borissova et al. 
1999; Stetson et al. 1999; and VandenBerg 1999a for discussions 
of the metallicities of M5, Pal~4, and Eridanus). Consistent 
with our working hypothesis that age is the sole  
second parameter, we assume that M5 and Pal~4/Eridanus have the 
same chemical composition. 

We have assumed throughout this paper that the HB morphology 
of the studied GCs can be reproduced by unimodal Gaussian 
deviates in ZAHB mass (see Catelan et al. 1998 for a detailed 
discussion).

A relevant 
numerical improvement is the adoption of Hill's (1982) 
interpolation algorithm also to interpolate among the evolutionary 
tracks of different masses in order to infer the physical parameters 
$\log\,L$, $\log\,T_{\rm eff}$ of the ``stars" in the HB simulations.
The synthetic HBs were converted to the observational planes using 
the prescriptions provided by VandenBerg (1999b). 

\subsection{The Case of M5}
Synthetic HBs have been computed aiming at estimating the
optimum parameters $\langle M_{\rm HB} \rangle$ (mean mass)
and $\sigma_M$ (mass dispersion) required to reproduce the 
observed HB morphology parameters for M5 (Table~1). The 
adopted procedure is completely analogous to that 
employed by Catelan et al. (1998).

We have computed synthetic HBs assuming an
overall number of HB stars $B+V+R = 553$, as in Sandquist's
(1998) sample. For each 
($\langle M_{\rm HB} \rangle$,~$\sigma_M$) combination, we
computed a series of 100 Monte Carlo simulations and 
obtained HB morphology parameters therefrom. After many 
such trials varying the above two free parameters, we have
converged on a set of models characterized by the following
values:

\begin{displaymath}
      \langle M_{\rm HB} \rangle = 0.6325\, M_{\sun},\,\,\,
                       \sigma_M = 0.025\, M_{\sun}. 
\end{displaymath}

\noindent Such a combination leads to the mean HB morphology
parameters described in Table~2 (where the numbers in parentheses 
represent the standard deviation of the mean over the set of 100 
simulations with 553 ``stars" in each). Note the nice 
agreement between the observed (Table~1) and theoretical 
(Table~2) parameters, to within the errors.

It should be remarked that, if Buonanno's (1993) parameter were 
bluer (implying a higher reddening; cf. \S2.1), our simulations
indicate that it would have been easier to account for the 
overall ratio between blue stars and RR Lyrae variables. Indeed, 
Schlegel et al. (1998) give $E(\bv) = 0.038$~mag for this 
cluster. The $\langle M_{\rm HB} \rangle$ value would not
differ significantly from the one quoted above though.

\vskip 0.25in
\parbox{3in}{\epsfxsize=3.25in \epsfysize=1.35in \epsfbox{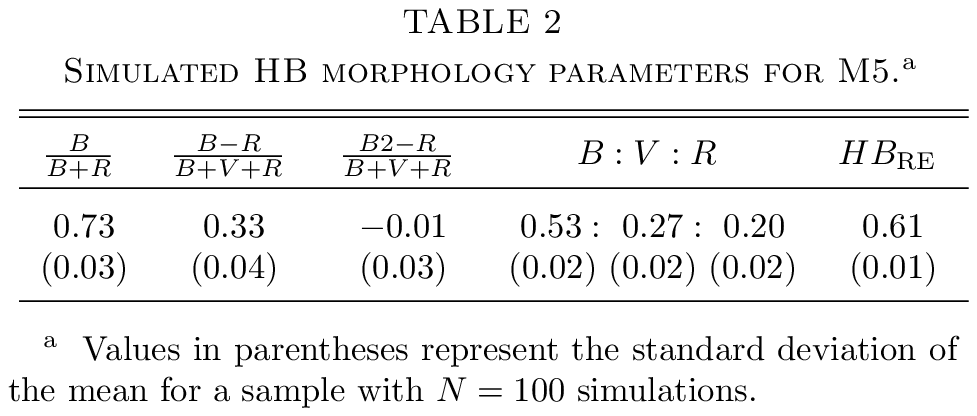}} 
\vskip 0.25in

It follows that the above $\langle M_{\rm HB} \rangle$ value 
for M5 is a quite robust result for the assumed chemical
composition and theoretical framework. In Figure~2, we
plot two synthetic HB/upper RGB models for M5, picked at 
random from the pool of 100 
simulations. The plus signs indicate RR Lyrae variables (a strip 
width of 0.075 in $\log\,T_{\rm eff}$ has been assumed). Random 
scatter has been included following the prescriptions of 
Robertson (1974), but without any special effort to make the 
CMD dispersion on the RGB match closely the observed one in 
Sandquist et al. (1996). 

\subsection{The Case of Pal~4/Eridanus}
To model the HBs of the extreme outer-halo clusters Pal~4 and
Eridanus is a significantly more complicated and challenging
task than to model that of M5, given the small number of HB 
stars detected in the HST studies and the total lack of HB stars 
lying blueward of the red HB. 

After trying a few different possibilities, we decided to adopt 
the following approach. Starting with a mass distribution which, 
in the mean, would give roughly the same number of stars on the 
red HB and inside the instability strip, we ran many {\em sets} 
of (twelve) synthetic HB simulations, increasing the mean mass by 
0.01~$M_{\sun}$ from one set to the next, and holding the mass 
dispersion (as well as the total number of HB stars---25) fixed 
(at $\sigma_M = 0.01 \, M_{\sun}$) in all cases. The mean mass 
range covered by our simulations was the following: 
$\langle M_{\rm HB}\rangle = 
 0.65,\,0.66\,\ldots\,0.78,\,0.79\,\,M_{\sun}$.

Again, random scatter was added following Robertson 
(1974). Here, however, we did make an effort to 
reproduce (approximately) the errors in the HST photometry
(Stetson 1999) around the HB level. Upon inspection of each
of the plots thus produced, and paying particular attention
to their corresponding color distributions in comparison to 
that shown in Figure~1, we reached the conclusion that the 
following parameters provide an adequate match to both the 
Pal~4 and the Eridanus HST CMDs at the HB level:

\begin{displaymath}
      \langle M_{\rm HB} \rangle = 0.75\, M_{\sun},\,\,\,
                       \sigma_M = 0.01\, M_{\sun}.
\end{displaymath}

\noindent Plots containing the simulations for this case can be 
found in Figure~3. The vertical dotted lines, as in Figure~1, 
indicate the color $\langle (V-I)_0 \rangle = 0.78$~mag. Notice 
that in some cases even the ``bright" red HB stars found 
(especially) in the Eridanus HST CMD are well reproduced. We 
interpret these stars as being the result of evolution away 
from the ZAHB towards the asymptotic giant branch, combined 
with statistical fluctuations due to the small sample size.

\parbox{3in}{\epsfxsize=3.0in \epsfysize=5.0in \epsfbox{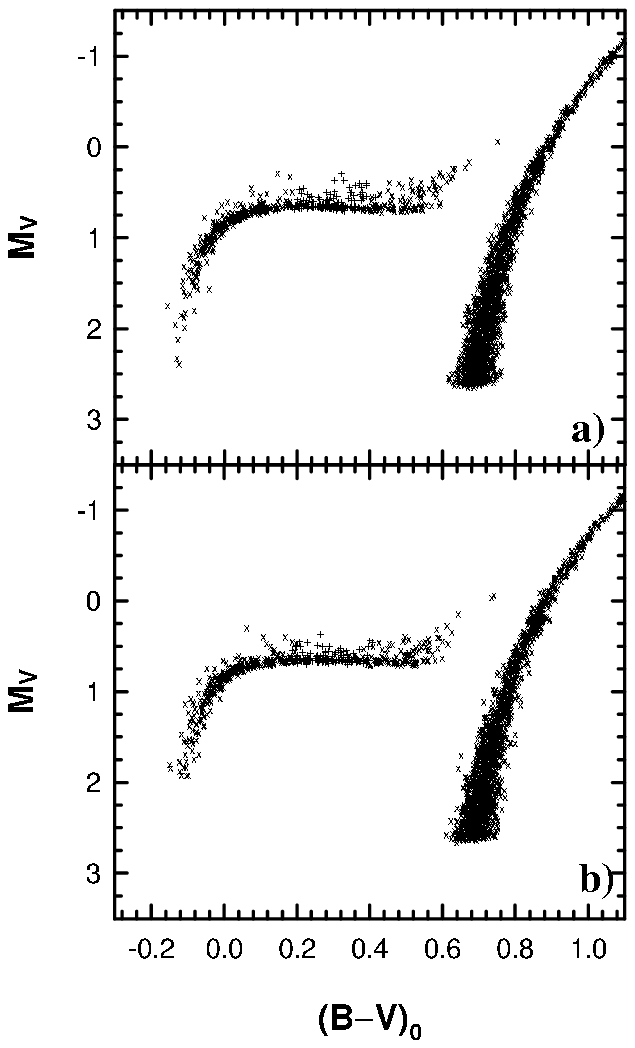}} 
\vskip 0.1in
\centerline{\parbox{3.5in}{\footnotesize  {\sc Fig.~2.---}
 Synthetic CMDs for M5 (see text).
 }}
\vskip 0.25in

%
\begin{figure*}[th]
 \figurenum{3} 
 \centerline{\epsfig{file=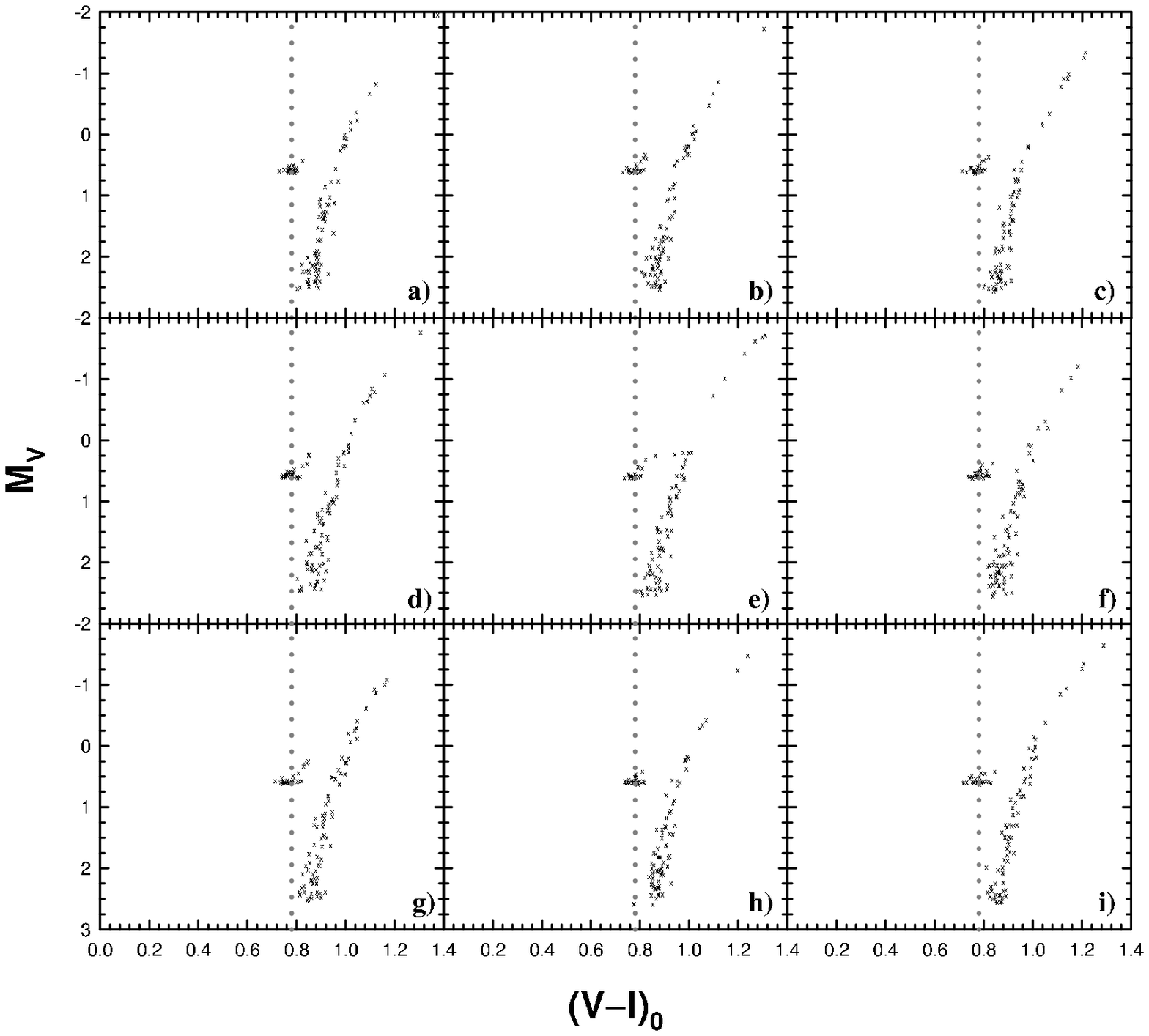,height=5.25in,width=5.75in}}
 \caption{Synthetic CMDs for Pal~4/Eridanus. 
         The distributions were obtained from Gaussian deviates with 
         $\langle M_{\rm HB} \rangle = 0.75\, M_{\sun}$, 
         $\sigma_M = 0.01\,M_{\sun}$, which we consider to be best 
         estimates. The vertical dotted lines 
         (in gray) indicate the $\langle (V-I)_0 \rangle = 0.78$~mag 
         locus (see Fig.~1).
         }
\end{figure*}
%
\begin{figure*}[t]
 \figurenum{4}
 \centerline{\epsfig{file=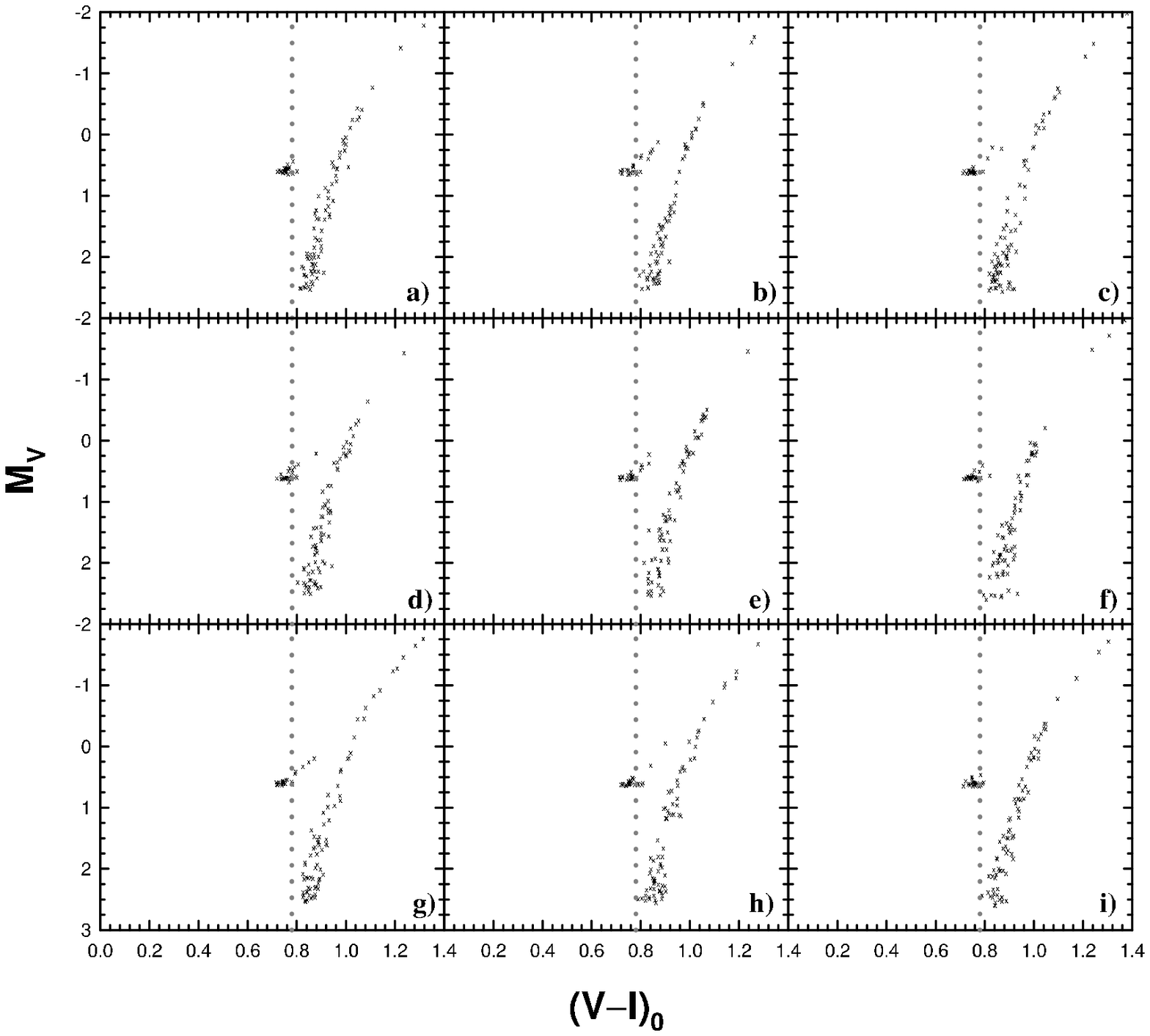,height=5.25in,width=5.75in}}
 \caption{As in Figure~3, but assuming that 
         $\langle M_{\rm HB} \rangle = 0.73\, M_{\sun}$.
         }
\end{figure*}

\vskip 0.25in
\begin{figure*}[hb]
 \centerline{\epsfig{file=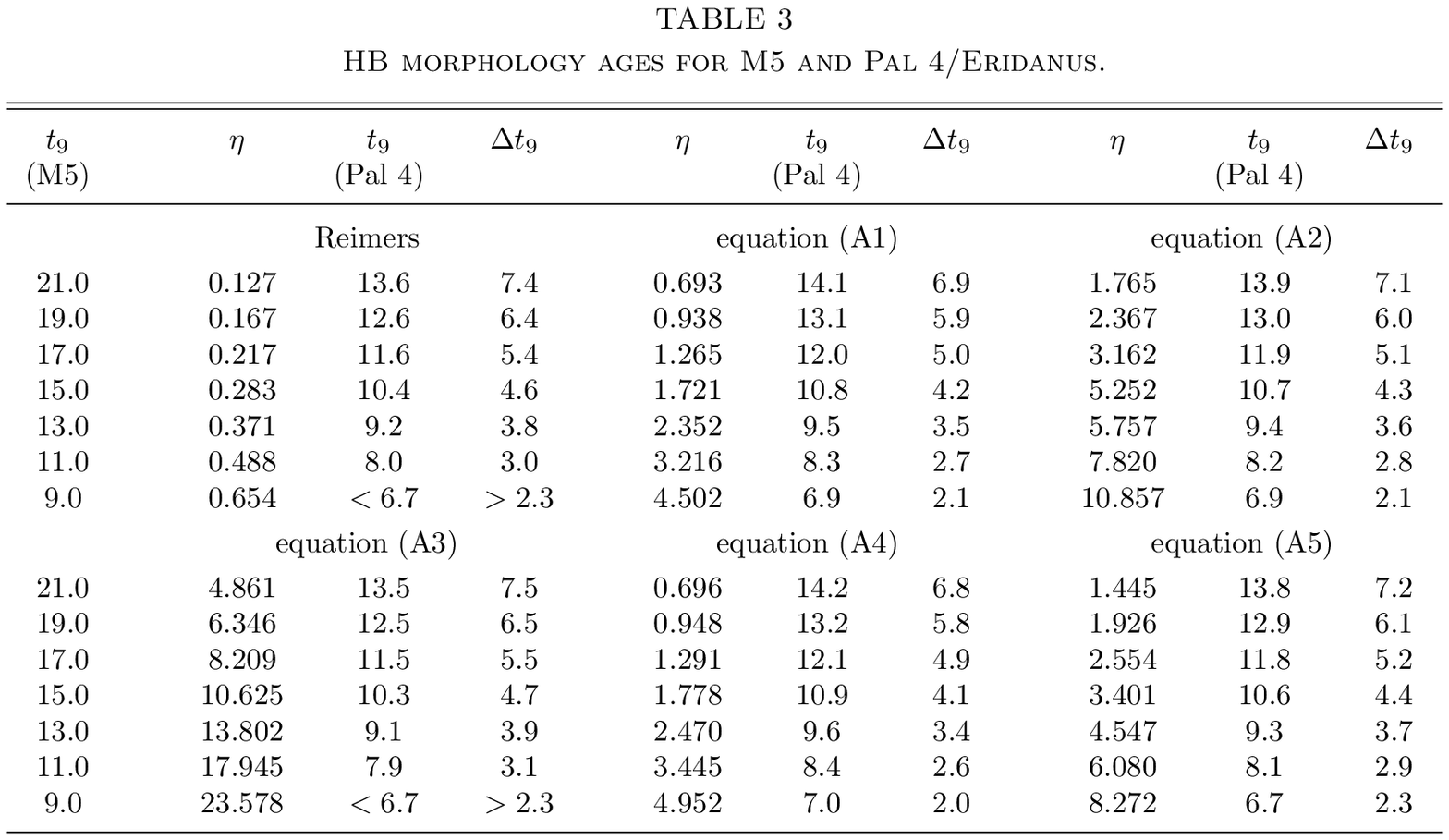,height=3.25in,width=5.35in}}
\end{figure*}
\vskip 0.25in

The reddening values from Schlegel et al. (1998) are generally 
larger than the commonly employed values tabulated by Harris 
(1996). In the case of Pal~4, one has: 

\begin{displaymath}
 E(\bv) = 0.023~{\rm mag} \Rightarrow E(V-I) \simeq 0.030~{\rm mag} 
\end{displaymath}

\noindent from Schlegel et al. (1998); and 

\begin{displaymath}
 E(\bv) = 0.01~{\rm mag} \Rightarrow E(V-I) \simeq 0.013~{\rm mag} 
\end{displaymath}

\noindent from Harris (1996). In the case of Eridanus, one finds: 

\begin{displaymath}
 E(\bv) = 0.022~{\rm mag} \Rightarrow E(V-I) \simeq 0.029~{\rm mag} 
\end{displaymath}

\noindent from Schlegel et al. (1998); and 

\begin{displaymath}
 E(\bv) = 0.02~{\rm mag} \Rightarrow E(V-I) \simeq 0.026~{\rm mag} 
\end{displaymath}

\noindent from Harris (1996). 

The transformation between $E(\bv)$ 
and $E(V-I)$ was carried out adopting a ratio of $\approx 1.3$ 
between the two (see Stetson et al. 1999). As is apparent, the 
difference in $E(V-I)$ values between the two sources is smaller 
in the case of Eridanus.    

This uncertainty in the $E(V-I)$ value for Pal~4 may affect the 
choice of model for the red-HB clusters. The larger reddening 
from Schlegel et al. (1998) implies an intrinsically {\em bluer} 
HB distribution than what would be inferred from Harris (1996).  

Because of possible systematic uncertainties in $E(V-I)$ and in 
the color transformations, we show, in Figures~4 and 5, synthetic 
HBs similar to those displayed in Figure~3, but varying 
$\langle M_{\rm HB} \rangle$ by $-0.02\,M_{\sun}$ (Fig.~4) and 
$+0.02\,M_{\sun}$ (Fig.~5). The latter case might be considered 
more appropriate if the canonical reddening value from Harris 
(1996) were adopted---implying larger relative ages with respect 
to M5 (see below). 

%
\begin{figure*}[th]
 \figurenum{5}
 \epsscale{0.5}
 \centerline{\epsfig{file=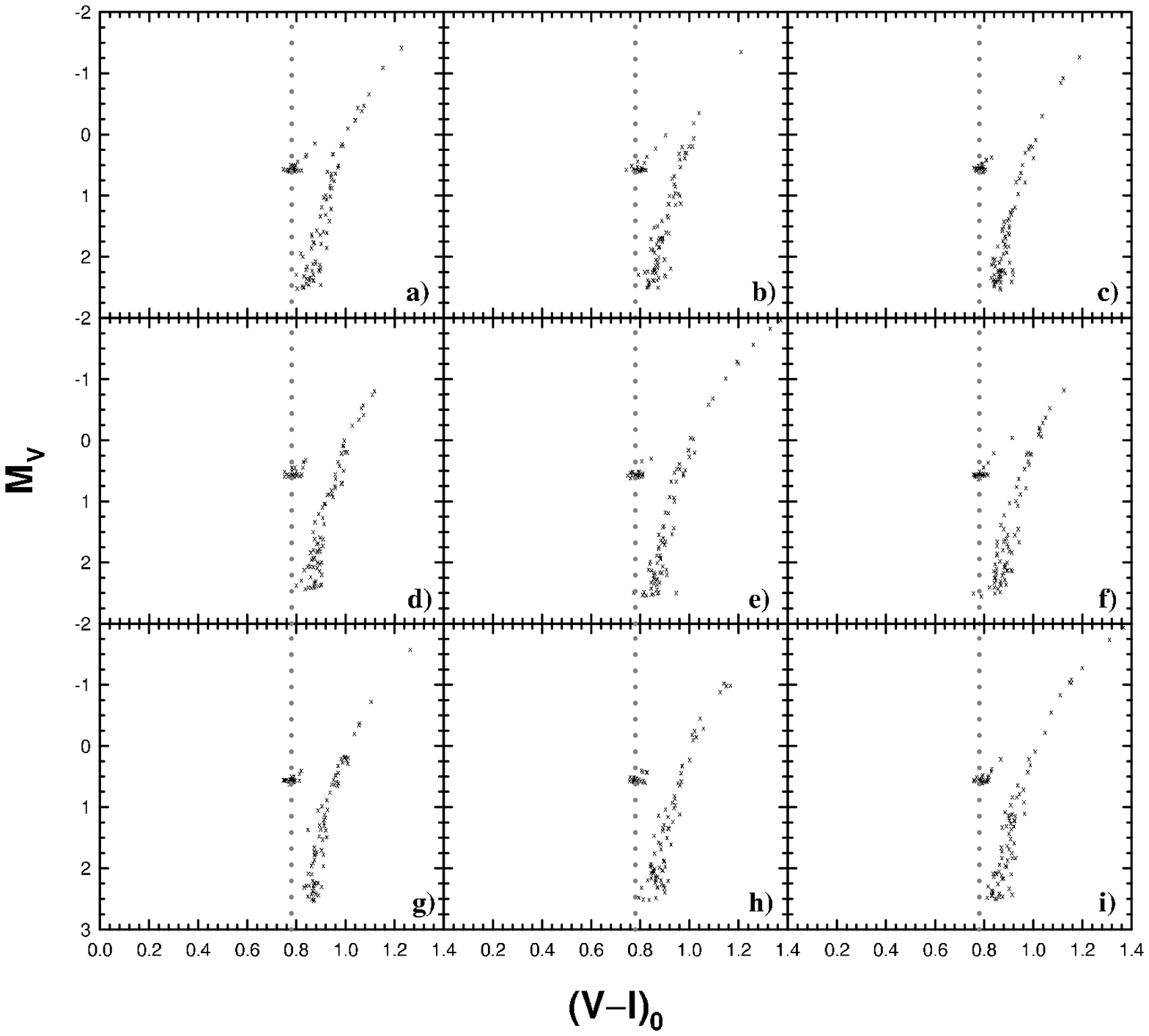,height=5.25in,width=5.75in}}
 \caption{As in Figure~3, but assuming that 
         $\langle M_{\rm HB} \rangle = 0.77\, M_{\sun}$. These  
         provide a better representation of Pal~4/Eridanus, in  
         case the reddening values from Harris (1996) are assumed
         instead of those from Schlegel et al. (1998).  
         }
\end{figure*}

Inspection of Figures~4 and 5 shows 
that statistical fluctuations can also play a role---though it 
appears more likely that this would lead to a larger  
$\langle M_{\rm HB} \rangle$ for Pal~4/Eridanus. Indeed, more 
of the models with larger $\langle M_{\rm HB} \rangle$ 
(Fig.~5) resemble those in Figure~3 in terms of  
$\langle (V-I)_0 \rangle$  than is the case for models with 
smaller $\langle M_{\rm HB} \rangle$ (Fig.~4). This is because of 
the decreasing dependence of HB temperature/color on stellar mass  
towards the red end of the HB. Thus, if important, statistical 
fluctuations would tend to lead to an underestimate of the age 
difference between M5 and Pal~4/Eridanus, as derived from HB 
morphology arguments.  

A set of 700 synthetic HB computations shows that changing 
$\sigma_M$ from $0.01\,M_{\sun}$ to $0.025\,M_{\sun}$ leads to 
a change in mean HB color equivalent to a reddening error by 
$\Delta\,E(\bv) \simeq 0.002$~mag, holding 
$\langle M_{\rm HB} \rangle$ fixed at $0.75\,M_{\sun}$. 
Had a larger, ``M5-like" 
$\sigma_M$ value been adopted, we would have been forced to adopt 
a slightly larger $\langle M_{\rm HB} \rangle$ for Pal~4/Eridanus. 
It is easy to see why: for a given $\langle M_{\rm HB} \rangle$, 
the low-mass tail of the distributions gets closer and closer to 
the instability strip region with increasing $\sigma_M$. This 
must be compensated for by increasing $\langle M_{\rm HB} \rangle$, 
also implying a (slightly) larger age difference between 
Pal~4/Eridanus and M5 than reported in the next sections.
However, we have been unable to obtain as satisfactory matches 
to the HBs of Pal~4/Eridanus using the larger $\sigma_M$, 
possibly pointing to a real difference in the mass dispersion 
among these systems.

\section{Estimating Relative Ages from the HB and RGB Models}

In order to estimate the relative ages required to produce the
relative HB types of M5 and Pal~4/Eridanus, we follow a similar 
approach as described in previous papers of this series (e.g.,
Catelan \& de Freitas Pacheco 1995). The main difference here 
is that we shall evaluate {\em the effects of an age-dependent 
mass loss on the RGB}, as implied by 
the several different analytical formulae discussed in the  
Appendix, upon the relative ages thus 
estimated. RGB mass loss is estimated on the 
basis of the RGB models of VandenBerg 
et al. (2000) for a chemical composition 
${\rm [Fe/H]} = -1.41$, $[\alpha/{\rm Fe}] = +0.3$. It is 
important to note that the VandenBerg et al. results for both 
the RGB and HB phases are in very good agreement with those 
from A.~V.~Sweigart (see VandenBerg et al. for a discussion). 

From the VandenBerg et al. (2000) models, we first obtained 
the age--RGB tip mass ($M_{\rm RGB}^{\rm tip}$) relationship 
for the adopted chemical composition. Then we obtained the 
age--overall mass loss on the RGB 
($\Delta\,M_{\rm RGB}^{\rm tip}$) relationship from the 
Appendix. The required estimate of $\eta$ values was then 
accomplished by evaluating, for each given (assumed) age for 
M5,

\begin{equation}
 \eta = 
 \frac{M_{\rm RGB}^{\rm tip} - M_{\rm HB}}{\Delta\,M_{\rm RGB}^{\rm tip}},  
\end{equation}

\noindent where $M_{\rm HB} \equiv 
   \langle M_{\rm HB} \rangle_{\rm M5} = 0.6325\,\,M_{\sun}$. 

Holding the $\eta$ value thus derived fixed, the age that 
leads to a good fit to the Pal~4/Eridanus HB morphology 
(as characterized by the mean HB mass value described in the 
previous section) was easily obtained from 
the $M_{\rm RGB}^{\rm tip}$--age and 
$\Delta\,M_{\rm RGB}^{\rm tip}$--age relationships. Hill's 
(1982) algorithm was used for the interpolations that defined 
such relationships. The implied age difference between M5 and 
Pal~4/Eridanus followed immediately from this. 

Table~3 shows our derived $\eta$ values and ages 
for Pal~4/Eridanus for each assumed age for M5 and for each  
of the mass loss formulae discussed in the Appendix, 
including Reimers' (1975a, 1975b) widely adopted one. The 
inferred age differences are also listed. Figure~6 summarizes 
our results; the hatched regions indicate the relative ages 
favored by the HST analyses of Stetson et al. (1999) and 
VandenBerg (1999a). 

From Table~3 and Figure~6, it is clear that 
only for extremely low ages---$\lesssim 9$~Gyr---can one 
reproduce the relative HB types of M5 and Pal~4/Eridanus in 
terms of ``age as the second parameter," even by assuming 
several different possibilities for the form of an 
age-dependent mass loss formula for giant stars. 
Equation~(A2) is the one which leads to the smaller age 
differences from HB morphology arguments. Except for 
equation~(A3), Reimers' (1975a, 1975b) is the one from 
which the largest relative 
ages are inferred. Note that only the {\em upper limits} 
on the possible age difference range estimated by 
Stetson et al. (1999) are reached for M5 ages of about 
9~Gyr.\footnote{This result is actually somewhat 
underemphasized by the way 
we have chosen to present the Stetson et al. (1999) and 
VandenBerg (1999a) results in Figure~6. The reason for this 
is that, in these studies, an absolute age of $\sim 13-15$~Gyr 
was adopted, whereas an absolute age $\lesssim 10$~Gyr would 
lead to even smaller relative turnoff ages. In other words, 
what we show as perfectly horizontal hatched areas in 
Figure~6 should actually be somewhat slanted, with the 
relative turnoff ages decreasing with decreasing M5 
age---thereby making it even harder for the relative ages
derived from HB morphology arguments to match the relative 
turnoff ages obtained from deep HST photometry, even for 
very low absolute ages.} VandenBerg's (1999a) results are 
not reproduced at all; extrapolation of the curves shown 
in Figure~6 suggests that an M5 age $\lesssim 8$~Gyr would 
be required to match the relative ages derived by VandenBerg. 

The situation would become somewhat less critical if the 
synthetic HBs shown in Figure~4 were adopted for Pal~4/Eridanus, 
as shown in Figure~7. Absolute ages for M5 of $\lesssim 10$~Gyr 
would be required in this case, and it might be possible to 
achieve agreement with VandenBerg's (1999a) results for an 
M5 age of $\lesssim 9$ Gyr. However, the opposite holds if  
the synthetic HBs displayed in Figure~5 are adopted instead, 
as one can see from Figure~8. We recall, from the arguments 
in the previous section, that the models in Figure~5 might 
be the better alternative to those in Figure~3 as genuinely  
representing Pal~4/Eridanus, if one were to adopt the 
canonical reddening values from Harris (1996) or if 
statistical fluctuation effects are important.  

In summary, we conclude that the requirement of extremely low 
ages, $< 10$~Gyr, for all GCs under consideration is a 
very firm result of the present investigation.   

%
\begin{figure*}[t]
 \figurenum{6}
 \centerline{\psfig{figure=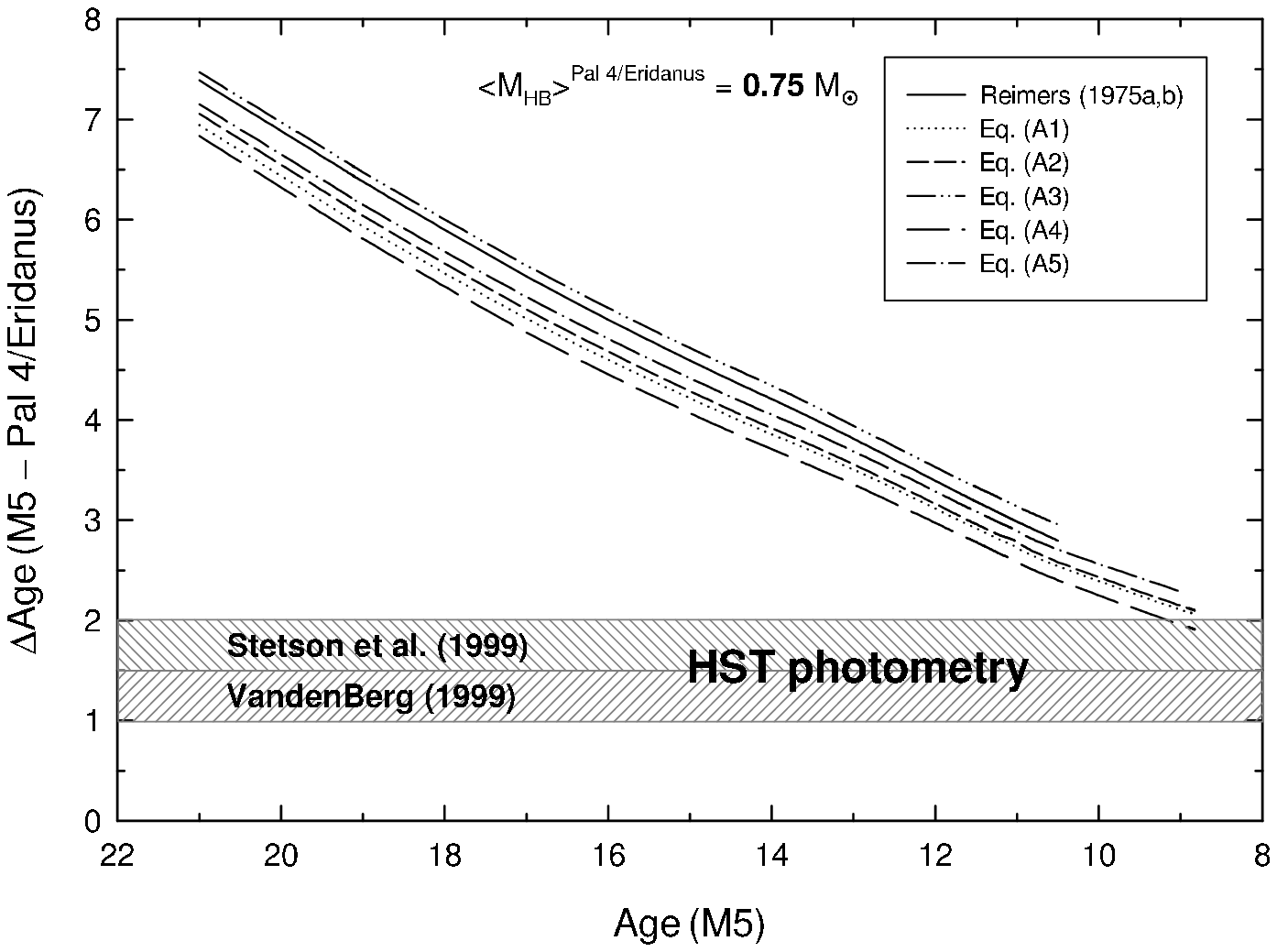}}
 \caption{The difference in age 
    $\Delta{\rm Age}({\rm M5-Pal~4/Eridanus})$ (in Gyr), 
    derived for the several indicated mass loss formulae,  
    is plotted as a function of the 
    M5 age (also in Gyr). This figure corresponds 
    to our preferred HB models for Pal~4/Eridanus, plotted 
    in Figure~3, obtained for   
    $\langle M_{\rm HB} \rangle = 0.75\,M_{\sun}$. 
    The hatched areas correspond to the range in turnoff  
    age differences between M5 and Pal~4/Eridanus, as 
    estimated by Stetson et al. (1999) and VandenBerg 
    (1999a) from deep HST photometry.
  }
\end{figure*}

\section{Conclusions and Discussion}
As far as the second-parameter phenomenon goes, we have 
demonstrated that age cannot be the {\em only} second 
parameter at play, unless one is willing to accept that  
the ages of M5-like GCs are less than 
10~Gyr.\footnote{One should bear in mind that this result 
depends critically on the accuracy of the HST-WFPC2 data
(see, e.g., Stetson 1998 in regard to charge-transfer 
effects), particularly the photometric zero points. 
Analysis of this subject is beyond the scope of this 
paper.} The same conclusion was reached in previous papers 
of this series (Catelan \& de Freitas Pacheco 1993, 1994, 
1995; see also Ferraro et al. 1997b) but, in the present 
study, we have fully taken into account the effects of an 
age-dependent mass loss on the RGB.  

If the [$\alpha$/Fe] ratio in Pal~4/Eridanus is 
lower than commonly found among GCs (Carney 1996), resembling 
instead the cases of the ``young," loose GCs Ruprecht~106 and 
Pal~12 (Brown, Wallerstein, \& Zucker 1997)---a hypothesis 
which is perhaps not unlikely, 
since all these clusters may have partaken a common origin 
(Majewski 1994; Fusi Pecci et al. 1995; Lynden-Bell \& 
Lynden-Bell 1995)---the ``turnoff age difference" between M5 
and Pal~4/Eridanus would {\em decrease}, as inferred from 
isochrone fits to the HST data (Stetson et al. 1999). On the 
other hand, the ``HB morphology age difference" would 
{\em increase} significantly, as inferred from the 
relative HB types of Pal~4/Eridanus vs. M5, because the 
red HBs of Pal~4/Eridanus would require a further decrease 
in the ages needed to match their HBs than derived in the
present paper, due to their lower overall [M/H].  

We note that proper-motion studies indicate that M5 
too is an {\em outer-halo} GC, which just happens to lie close to 
its perigalacticon (Cudworth 1997 and references therein). 
According to such work, M5 actually spends much of its time at 
galactocentric distances larger than $\sim 50$~kpc. Therefore, 
we should keep in mind
that, when using M5 to compare its age against those of (other) 
outer-halo GCs, we may simply be measuring the age dispersion
in the outer Galactic halo, and not the age difference between
the inner and the (extreme) outer halo---contrary to what is
often assumed.

\acknowledgments
The author wishes to express his gratitude to D.~A. VandenBerg 
for providing many useful comments and suggestions, and also 
for making his latest evolutionary computations available in 
advance of publication. F.~R. Ferraro, E.~L. Sandquist, and 
P.~B. Stetson have supplied crucial observational 
data and/or information, and are also warmly thanked. Useful 
comments by F.~Grundahl, W.~B.~Landsman, and R.~T.~Rood are 
gratefully acknowledged, as are the suggestions by an 
anonymous referee which greatly helped improve the 
presentation of these results. Support for this 
work was provided by NASA through Hubble Fellowship grant 
HF--01105.01--98A awarded by the Space Telescope Science 
Institute, which is operated by the Association of 
Universities for Research in Astronomy, Inc., for NASA under 
contract NAS~5--26555.

\appendix

\vspace{0.5cm}

\centerline{{\sc Analytical Mass Loss Formulae Revisited}}

\vspace{0.5cm}

Mass loss on the RGB is widely recognized as one of the 
most important ingredients, as far as the HB morphology goes (e.g., Catelan 
\& de Freitas Pacheco 1995; Lee et al. 1994; Rood, Whitney, \& D'Cruz 1997). 
Up to now, investigations of the impact of RGB mass loss upon the HB 
morphology have mostly relied on Reimers' (1975a, 1975b) mass loss formula. 
We note, however, that Reimers' is by no means the only mass loss formula 
available for this type of study. In particular, alternative formulae have 
been presented by Mullan (1978), Goldberg (1979), and Judge \& Stencel 
(1991, hereafter JS91). 

%
\begin{figure*}[t]
 \figurenum{7}
 \centerline{\psfig{figure=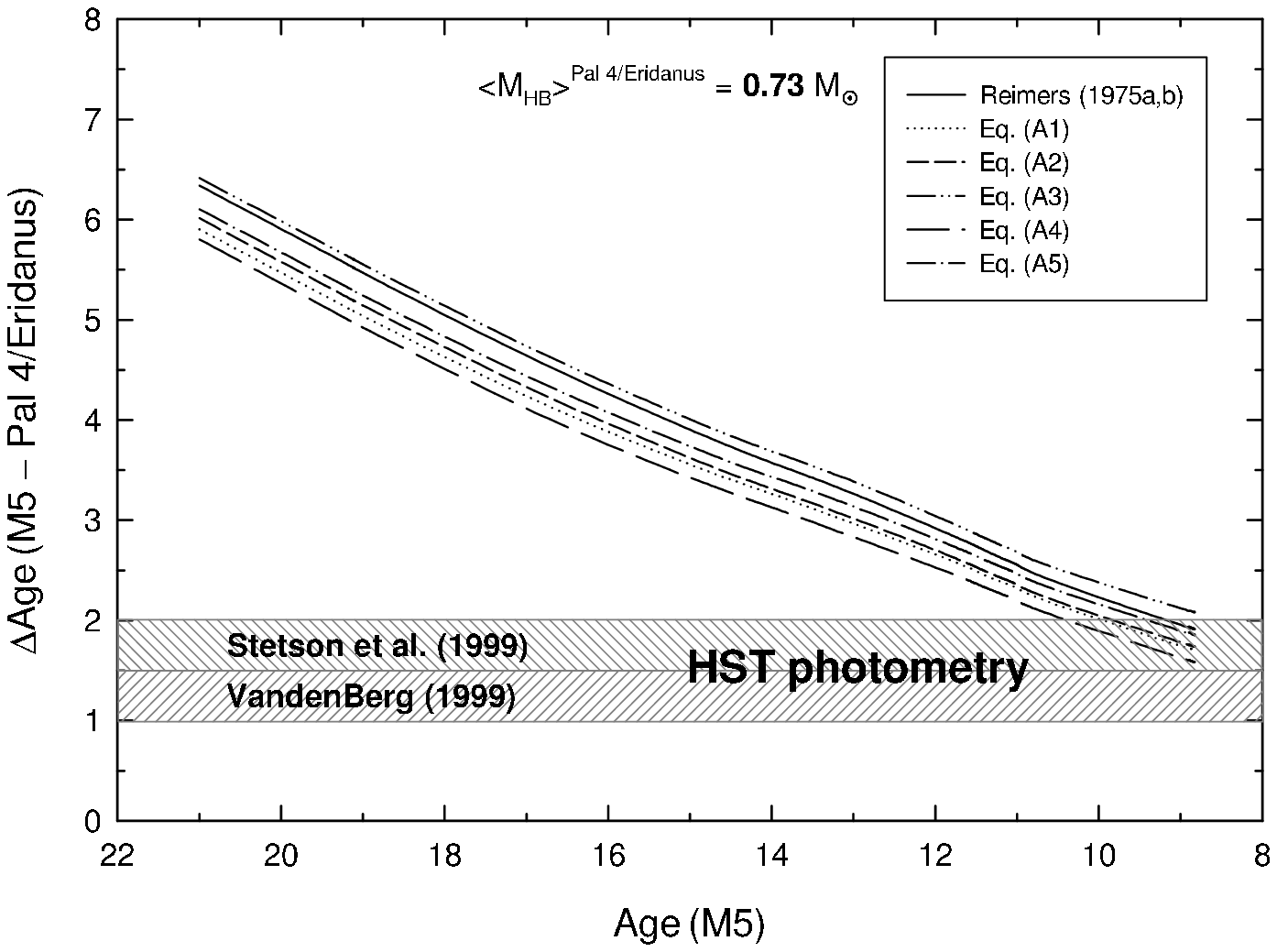}}
 \caption{As in Figure~6, but for the  
    models displayed in Figure~4, obtained for   
    $\langle M_{\rm HB} \rangle = 0.73\,M_{\sun}$. 
  }
\end{figure*}

We have undertaken a revision of all these formulae, employing the latest 
and most extensive dataset available in the literature---namely, that of 
JS91. The mass loss rates provided in JS91 were compared against more 
recent data, and excellent agreement was found (Fig.~A1). If the distance 
adopted by JS91 lied more than about $2 \sigma$ away from that based on 
{\sc Hipparcos} trigonometric parallaxes, the star was discarded. Only 
five stars (L$^2$~Pup, U~Hya, X~Her, g~Her, $\delta^2$~Lyr) 
turned out to be discrepant, in a sample containing more than 
20 giants. Employing ordinary least-squares (OLS) regressions and 
following the Isobe et al. (1990) guidelines [``if the problem is to 
predict the value of one variable from the measurement of another, then 
OLS($Y|X$) should be used, where $Y$ is the variable to be predicted"] we 
find that the following formulae provide adequate fits to the data (see 
also Fig.~A2):  

\begin{equation}
\eqnum{A1}
\frac{{\rm d}M}{{\rm d}t} = 8.5 \times 10^{-10} 
       \left(\!\!\!\begin{array}{c}
                L \\ \overline{g\,R}
            \end{array}\!\!\!\right)^{+1.4}\,\,
M_{\odot}\,{\rm yr}^{-1},
\end{equation}

\noindent with $g$ in cgs units, and $L$ and $R$ in solar units. As can  
be seen, this represents a ``generalized" form of Reimers' original mass 
loss formula, essentially reproducing a later result by Reimers (1987). 
Formally, the exponent (+1.4) differs from the one in Reimers' (1975a, 
1975b) formula (+1.0) at $\approx 3\sigma$;  

\begin{equation}
\eqnum{A2}
\frac{{\rm d}M}{{\rm d}t} = 2.4 \times 10^{-11} 
       \left(\!\!\!\begin{array}{c}
                g \\ \overline{R^{\frac{3}{2}}}
            \end{array}\!\!\!\right)^{-0.9}\,\,
M_{\odot}\,{\rm yr}^{-1},
\end{equation}

\noindent likewise, but in the case of Mullan's (1978) formula; 

\begin{equation}
\eqnum{A3}
\frac{{\rm d}M}{{\rm d}t} = 1.2 \times 10^{-15}  \, R^{+3.2}\,\,
M_{\odot}\,{\rm yr}^{-1},
\end{equation}

\noindent idem, Goldberg's (1979) formula. Interestingly, the 
exponent (+3.2) is indistinguishable from +3.0 to well within 
$1 \sigma$;  

%
\begin{figure*}[t]
 \figurenum{8}
 \centerline{\psfig{figure=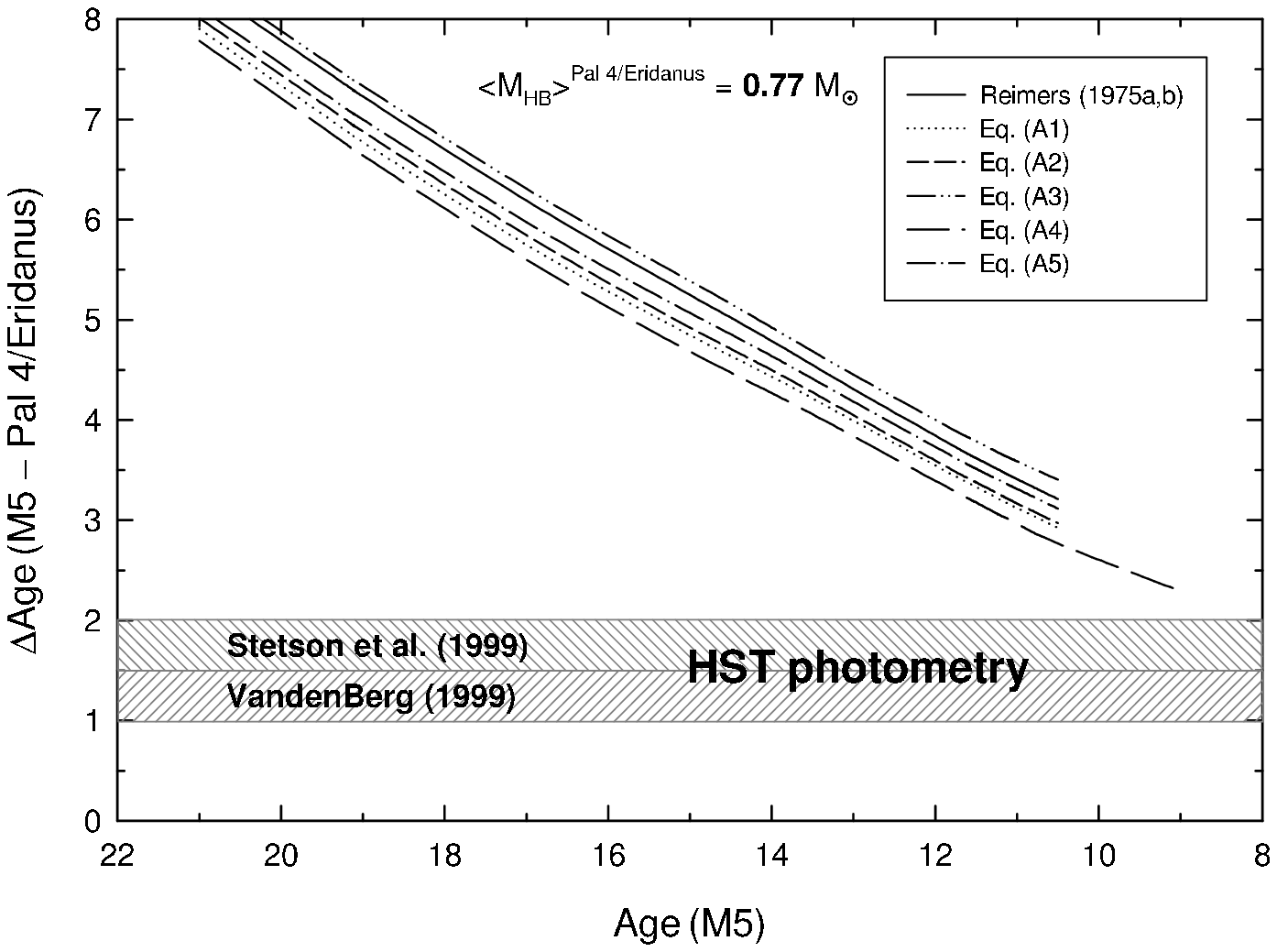}}
 \caption{As in Figure~6, but for the  
    models displayed in Figure~5, obtained for   
    $\langle M_{\rm HB} \rangle = 0.77\,M_{\sun}$. 
    This provides a better representation of Pal~4/Eridanus, in  
    case the reddening values from Harris (1996) are assumed 
    instead of those from Schlegel et al. (1998).
  }
\end{figure*}

\begin{equation}
\eqnum{A4}
\frac{{\rm d}M}{{\rm d}t} = 6.3 \times 10^{-8}  \, g^{-1.6}\,\,
M_{\odot}\,{\rm yr}^{-1},
\end{equation}

\noindent ibidem, JS91's formula. In addition, the expression 

\begin{equation}
\eqnum{A5}
\frac{{\rm d}M}{{\rm d}t} = 3.4 \times 10^{-12}  \, L^{+1.1}\, g^{-0.9}\,\,
M_{\odot}\,{\rm yr}^{-1}, 
\end{equation}

\noindent suggested to us by D.~VandenBerg, also provides a good 
fit to the data. ``Occam's razor"\footnote{
   ``{\em Entia non multiplicanda praeter necessitatem}." 
     (``Entities must not be multiplied beyond necessity.") 
     Occam's Razor is often referred to as the ``Principle of 
     Simplicity" or the ``Law of Parsimony" as well.} would 
favor equations~(A3) or (A4) in comparison with the others, but 
otherwise we are unable to identify any of them as being obviously 
superior. 

We emphasize that mass loss formulae such as those given above should 
not be employed in astrophysical applications (stellar evolution, 
analysis of integrated galactic spectra, etc.) without keeping in 
mind these exceedingly important limitations: 

1. As in Reimers' (1975a, 1975b) case, equations~(A1) through 
      (A5) were derived based on Population~I stars. Hence 
      they too are not well established for low-metallicity 
      stars. Moreover, there are only two first-ascent giants 
      ($\alpha$~Boo and $\beta$~Peg) in the adopted sample; 

2. Quoting Reimers (1977), ``besides the basic [stellar]
      parameters $\ldots$ the mass-loss process is probably also 
      influenced by the angular momentum, magnetic fields and close 
      companions. The {\em order of magnitude} of such effects is 
      completely unclear. Obviously, many observations will be
      necessary before we get a more detailed picture of stellar
      winds in red giants" (emphasis added). See also Dupree \&
      Reimers (1987); 

3. Similarly, Reimers (1975a) has pointed out that such mass 
      loss relations ``should be considered as interpolation 
      formulae only and not as strictly valid. Deviations due to 
      various other properties can be expected and must be left 
      to future research"; 

4. ``One should always bear in mind that a simple~$\ldots$ 
      formula like that proposed can be expected to yield only 
      correct order-of-magnitude results if extrapolated to the 
      short-lived evolutionary phases near the tips of the giant 
      branches" (Kudritzki \& Reimers 1978);

5. {\em Intrinsic} scatter among mass loss rates on the RGB 
      is expected to be present (e.g., Dupree \& Reimers 1987; 
      Rood et al. 1997; and references therein). The origin of 
      such scatter, inferred from the CMDs of GCs (e.g., Rood 
      1973; Renzini \& Fusi Pecci 1988), is currently unknown; 

%
\begin{figure*}[t]
 \figurenum{A1}
 \centerline{\psfig{figure=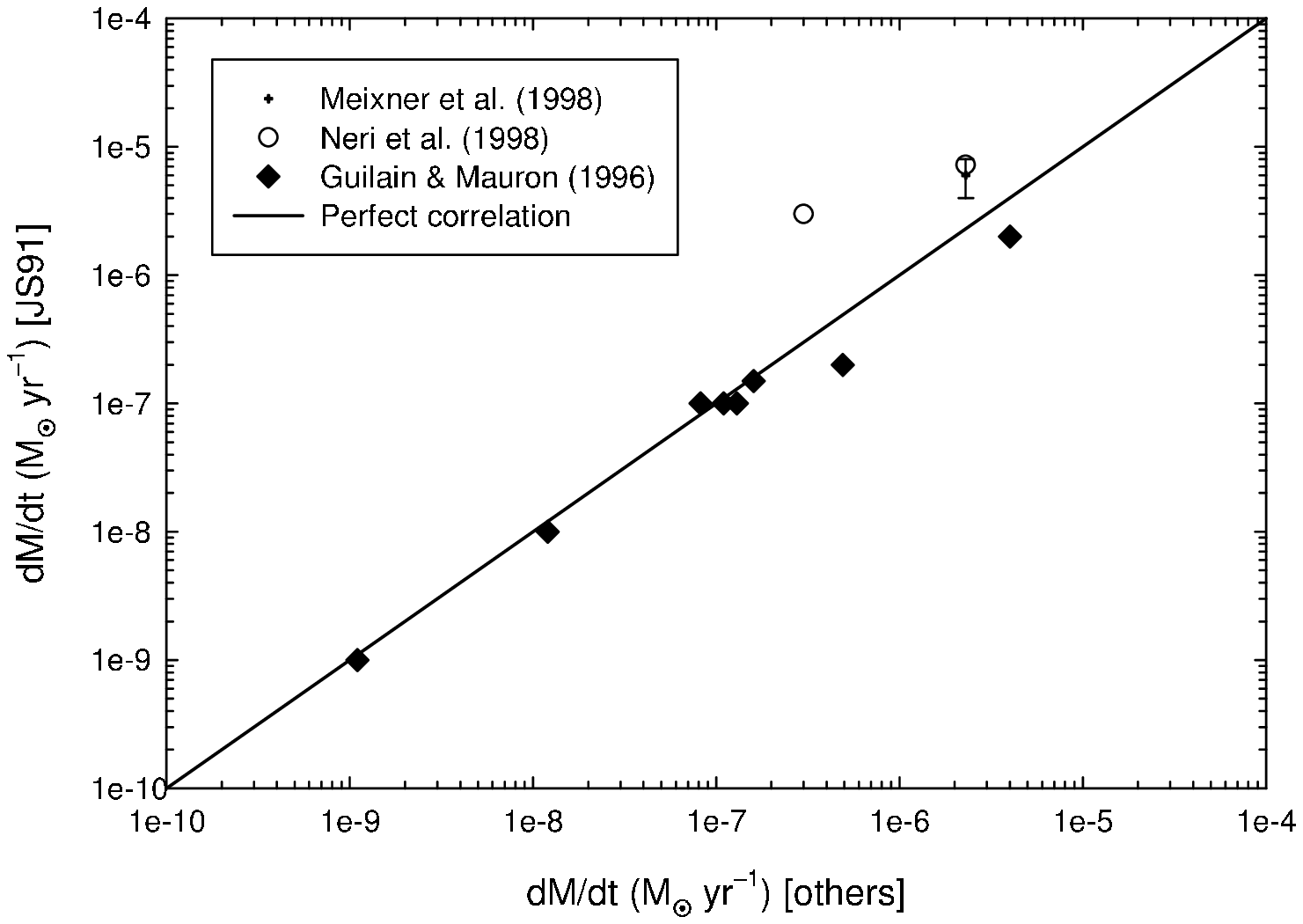}}
 \caption{Comparison between mass loss rates recently 
          provided in the literature with those tabulated by JS91. 
          There is clearly excellent overall agreement.   
          }
\end{figure*}

6. According to Willson (1999), ``correlations between observed  
      mass loss rates and physical parameters of cool stars may  
      be (and usually are) dominated by selection effects. Most 
      observations have been interpreted using models that are 
      relatively simple (stationary, polytropic, spherically 
      symmetric, homogeneous) and thus `observed' mass loss rates 
      or limits may be in error by orders of magnitude in some 
      cases." She further claims that ``Reimers' relation tells 
      us the properties of stars that are losing mass, and 
      {\em not} the mass loss rate that arises from a certain 
      set of stellar parameters. It has been widely misunderstood 
      and widely misused in stellar evolution and stellar 
      population studies";   

7. The two first-ascent giants analyzed by 
      Robinson, Carpenter, \& Brown (1998) using HST-GHRS, 
      $\alpha$~Tau and 
      $\gamma$~Dra, appear to both lie about one order of magnitude 
      below the relations that best fit the JS91 data---two orders 
      of magnitude in fact, if compared to Reimers' formula (see 
      Fig.~A2). The K supergiant $\lambda$~Vel, analyzed by the same 
      group (Mullan, Carpenter, \& Robinson 1998), appears in much 
      better agreement with the adopted dataset and best fitting 
      relations. 

{\em In effect, mass loss on the RGB is an excellent, but virtually 
untested, second-parameter candidate.} It may be connected to GC  
density, rotational velocities, and abundance anomalies on the RGB. 
It will be extremely important to study mass loss in first-ascent, 
low-metallicity giants---{\em in the field and in GCs alike}---using 
the most adequate ground- and space-based facilities available, or 
expected to become available, in the course of the next decade. 
Moreover, in order to properly determine how (mean) mass loss 
behaves as a function of the fundamental physical parameters and 
metallicity, astrometric missions much more accurate than 
{\sc Hipparcos}, such as SIM and GAIA, will certainly be necessary.  

{\em In the meantime, we suggest that using several different mass 
loss formulae constitutes a better approach than relying on a single 
one.}  

%
\begin{figure*}[t]
 \centerline{\psfig{figure=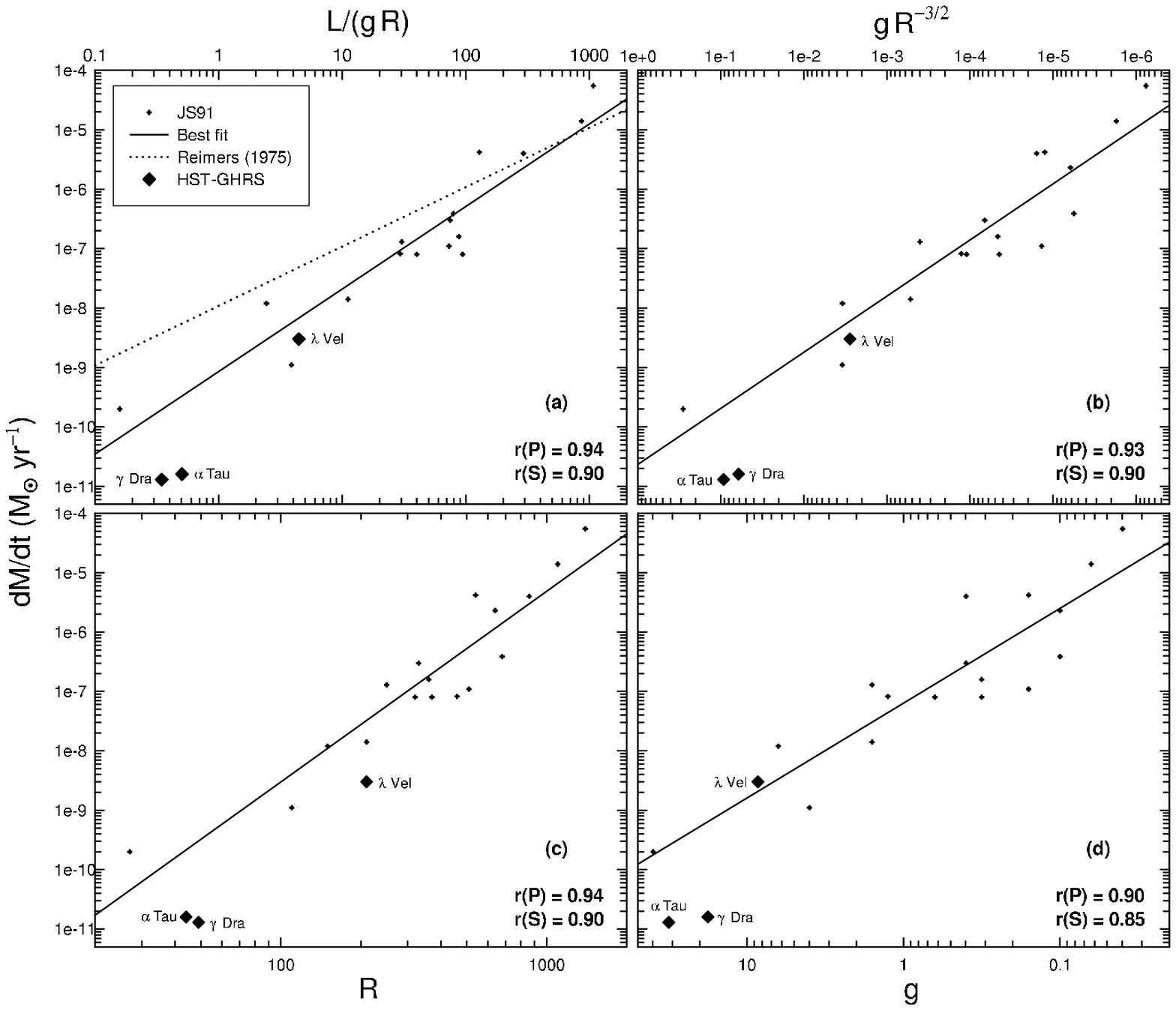}}
 \figurenum{A2}
\caption{Mass loss rate for cool giant stars (JS91)
     is plotted against $L/(g\,R)$
     (panel a), $g/R^{\frac{3}{2}}$ (b), $R$ (c), 
     and $g$ (d). All quantities are given in solar units except 
     for gravity which is in cgs units. 
     The continuous lines represent least-squares fits to the data
     [eqs.~(A1)--(A4)]. The Pearson and Spearman correlation 
     coefficients are given. HST-GHRS results are also shown.}
\end{figure*}

In this sense, the latest RGB evolutionary tracks by VandenBerg et al. 
(2000) were employed in an investigation of the amount of mass lost 
on the RGB and its dependence on age. As in some previous work (e.g., 
D'Cruz et al. 1996), the effects of mass loss upon RGB evolution 
were ignored, which is a good approximation except for those stars 
which lose a considerable fraction of their mass during their 
evolution up the RGB (e.g., Castellani \& Castellani 1993). 

In Figure~A3, the mass loss--age relationship is shown for each of 
equations~(A1) through (A5), and also for Reimers' (1975a, 1975b) 
formula, for a metallicity ${\rm [Fe/H]} = -1.41$, 
$[\alpha/{\rm Fe}] = +0.30$. {\em Note that even though these 
formulae are all based on the very same dataset} (JS91), {\em the 
implications do differ from case to case}.  

%
\begin{figure*}[t]
 \centerline{\psfig{figure=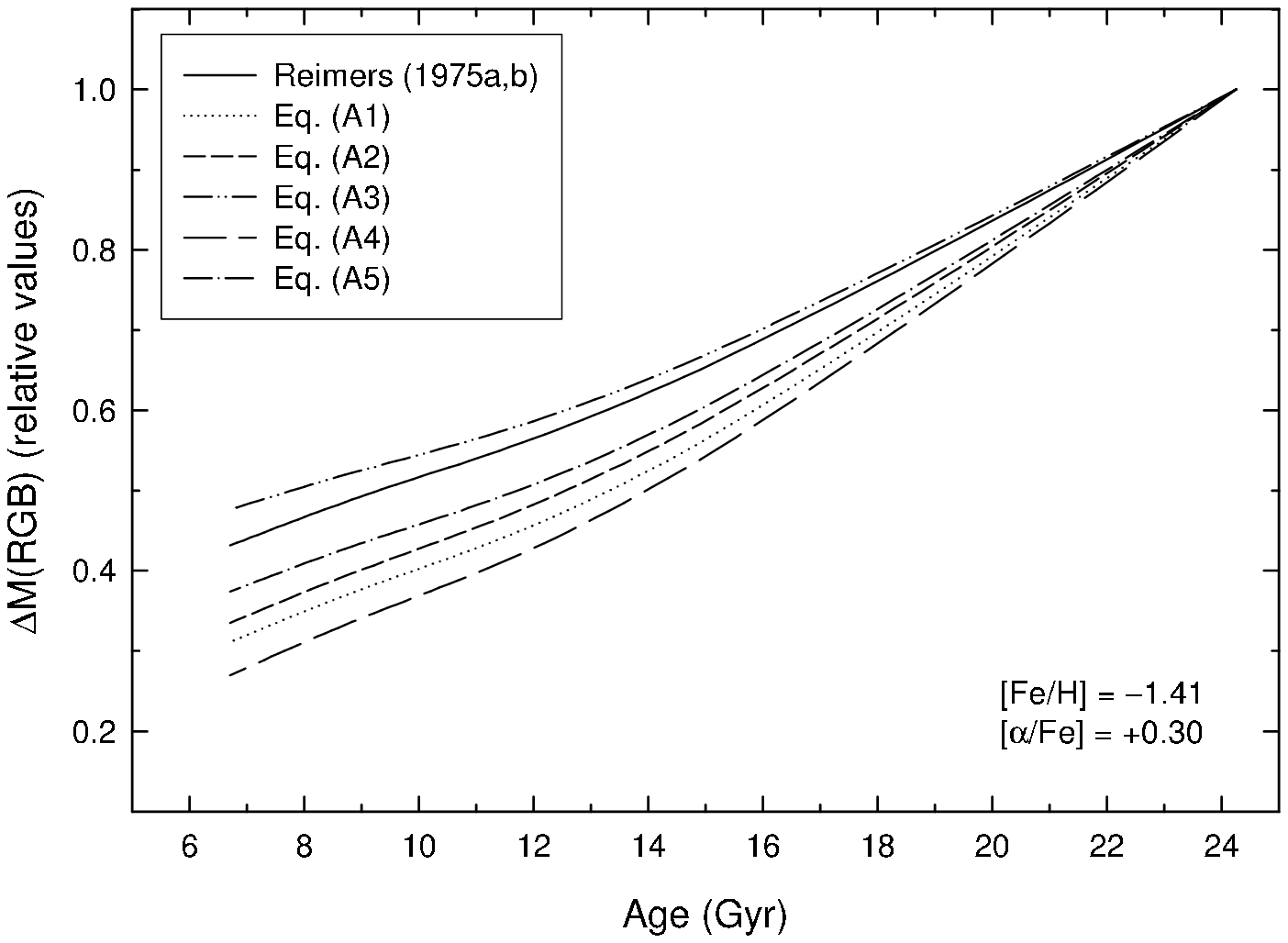}}
 \figurenum{A3}
\caption{Dependence of mass loss on age, for the chemical composition 
indicated at the lower right-hand corner. For each mass loss formula, 
mass loss values were normalized to the highest value, attained at an
age of 24.3~Gyr.}
\end{figure*}

\end{document}